# Self-organization phenomena at exciton condensation in quantum wells in an inhomogeneous external potential

**V.I. Sugakov**

Institute for Nuclear Research, NAS of Ukraine 47, Nauki Ave., Kyiv 03680, Ukraine

**Abstract**. A theory of the separation of a system of indirect excitons into a condensed and a gaseous phases with the formation of regular patterns of alternating phases in inhomogeneous external fields is developed. The theory is applied to the study of the non-uniform distribution of the exciton density in a double quantum well under a slot cut in a metallic electrode. It is shown that in a certain range of exciton generation rates a chain of light emitting islands periodically localized along the slot is developed. By creating a biased external potential along the slot the periodical pattern could be forced to move along the slot. Also the structures of condensed phases distribution arising at periodical external potential is considered.

**1.Introduction**

To present day the exciton condensed phase was not observed in bulk materials in spite of many efforts. Investigations of last years showed that indirect excitons in double quantum wells are perspective object for such condensation. Electrons and holes of indirect excitons are localized in different wells. A model of indirect excitons was suggested in [1]. Simultaneously and independently the exciton state with the localization of electron and hole in different layers was propozed in [2] for layer semiconductors, in which the interaction between layers is small. Moreover it was shown in the paper [3] that there is giant linear Stark effect for such type excitons because an electric field forces electrons and holes to different wells. As a result the indirect excitons are particles with large lifetimes due to the small overlap the wave functions. This fact allows achieving exciton systems with great density suitable for studying processes of the exciton-exciton interaction. Recent studies of the light emission of indirect excitons from semiconductor double quantum wells have reported the observation of the development of bright circles and spots often in regular patterns. The observed spatial structuring manifested itself as an appearance of inhomogeneities in the space distribution of the exciton density. Neither the inhomogeneity of the system nor any external forces can correspond to the regularity and symmetry of the spatial arrangement of the arising bright regions. The break of the uniform symmetry is spontaneous. For example, the authors of papers [4,5] observed luminescent rings very far from the exciting laser spot, at the distances significantly exceeding the exciton diffusion length. In some cases, the ring broke down into a number of periodically arranged fragments [4] in spite of the fact that the rate of the exciton generation caused by the electron-hole recombination on the ring was uniform everywhere along the ring's circumference. Authors of the paper [6] excited a double quantum well by light and measured excitonic emission from the wells through a circular window in a metallic electrode. They observed bright luminescent spots situated periodically on a circle under the rim of the window. Again, the conditions in different points under the rim were the same and, therefore, the observed structuring involved a symmetry break.

There are two approaches for the explanation of the development of the complex emitting patterns. In the fist group of works the explanations of the origin of the luminescent structures are based on the Bose-Einstein statistics applied to the systems of excitons [7-10] (on the Bose-Einstein condensation or on the processes in which Bose-Einstein statistics plays an important role). But these theories did not progress to the point of being able to perform detailed studies of the observed features, relate the theoretical parameters with the experimental data or describe the evolution of the patterns with the change of temperature, pumping intensity and the parameters of the exciton system.

The second approach was formulated in the papers [11-17]. It was shown in [11] that in the case of the attractive interaction between excitons the uniform destribution of the exciton density becomes unstable with respect to periodical structure development at a certain critical value of the pumping. The theory [14-17] explained the development of the patterns experimentally observed in Refs. [4,6] and also the other results, obtained in [18] for narrow line caused by condensed phase emission, such as the phase diagram - critical pumping vs. temperature, the dependence of luminescence intensity on temperature at fixed pumping and the dependence of luminescence on pumping at fixed temperature. The theory relied on two main assumptions:

1. There exists a certain condensed phase of excitons, caused by some short-range attractive interaction between indirect excitons. This attraction dominates over the dipole-dipole repulsion if the distance between the quantum wells is not very large. The attraction occurs mainly due to the exchange interaction between the electrons of the indirect excitons. The plausibility of such attractive interaction between indirect excitons finds support in recent theoretical studies. The theoretical possibility of the existence of an excitonic liquid phase in double quantum well structures has been shown in the paper] [19]. Additionally, the papers [20-22] have demonstrated that the indirect excitons may couple to form biexcitons.

2. Non-equilibrium state of system and , particularly, the finite value of exciton lifetime play an important role in the structure formation. plays an important role in the formation of the structures. Typically the exciton lifetime is much larger than the time of an establishment of local equilibrium, but it is less than the time of an establishment of an equilibrium between different phases, as the latter is controlled by slow diffusion processes.. The value of the lifetime determines the structures that arise in two phase's system. To study the system at parameters at which two phases coexistent taking into account of the finite value of lifetime is necessary. So, the structures are result of self-organization processes in non-equilibrium systems.

To present time there are different suggestions of physical systems for exciton trapping: laser induced trapping [23], trapping in harmonic potential [24,25]

There are two most popular mathematical models for the description of the growth of new phases during phase transitions that could be applied to the formation of spatial patterns of condensed and gaseous phases: the model of the nucleation and the model of the spinodal decomposition. Both were generalized by us for the case of unstable particles and used in the papers [14-17]. In these previous papers we considered systems which had already been studied experimentally. In the present paper we propose a new effect, yet to be experimentally discovered. We study the distribution of the density of the double quantum well indirect excitons and its behavior in a non-homogeneous external field created by a presence of a slot cut in a metallic electrode. Additionally, we predict the migration of the patterns along the slot in a non-homogeneous potential. Such motion is similar to the Gunn effect, in which a charge distribution moves along a semiconductor in an external electric field. The comparison of the theory and the experiment would produce an additional verification of the theory and present new effects in the behavior of the high density excitons forming a condensed phase.

**2. The main equation for the exciton density**

We consider a system in which high exciton density leads to the formation of regions of the condensed phase. The distribution of the exciton density will be studied in the framework of the model of the spinodal decomposition. In the considered system, where multiple regions of different phases are possible, three characteristic time parameters can be introduced: the time $t_l$ during which the local equilibrium is reached, the exciton lifetime $\tau$ and the time $t_m$ required for establishing the equilibrium between different islands of the condensed phase. Usually the condition $t_l << \tau$ holds and the system quickly reaches the local equilibrium. It is the validity of this criterion and the establishment of the local equilibrium that allow the treatment of excitons in the limit of $\tau \to \infty$, which is the typical approximation in many papers where a single phase states are considered. But the time of the establishment of the equilibrium between different regions of the condensed and the gaseous phases is large ($t_m \geq \tau$) because it is controlled by slow diffusion processes. This equilibrium is not reached during the exciton lifetime. Therefore, the exciton lifetime is an important parameter for the description the system in conditions of the coexistence of several phases.

We shall deduce the equation for the exciton density distribution using the exciton conservation law and phenomenological expressions of the non-equilibrium thermodynamics. The conservation law for the exciton density gives the following equation

$$\frac{\partial n}{\partial t} = -div\vec{j} + G(\vec{r}) - \frac{n}{\tau} \qquad (1)$$

where $G(\vec{r})$ is the exciton generation rate or pumping (the number of excitons created per unit area in unit time). The processes of the creation and emission of excitons may be described in such simple form if the lifetime of excitons is much larger than the time of the establishment of the local equilibrium in the well.

In general, the connection between the flux $\vec{j}$ and thermodynamical forces is non-local. In the case of uniform temperature this connection takes the form $\vec{j}(\vec{r},t) = -\int M(\vec{r}t,\vec{r}'t')\vec{\nabla}\mu(\vec{r}'t')d\vec{r}'dt'$ where $M(\vec{r}t,\vec{r}'t')$ is a phenomenological (kinetic) coefficient. It can be expressed via the flux correlation function $<\vec{j}(\vec{r}t)\vec{j}(\vec{r}'t')>$ [16] where averaging is carried out over the local equilibrium distribution. If the thermodynamical force $\vec{\nabla}\mu(\vec{r},t)$ changes slowly in time, compared to the characteristic duration of the damping of the correlation function, and in space, at the distances of the quantum correlation length, we may consider the connection between the exciton current and the gradient of the chemical potential $\mu$ to be local:

$$\vec{j} = -M\vec{\nabla}\mu \qquad (2)$$

where $M$ is the exciton mobility.

If the time of the establishment of the local equilibrium is significantly less than both the exciton lifetime and the time of the establishment of the equilibrium between various regions, the free energy of the quasi-local state can be considered as a function of the exciton density $n$. The chemical potential may be obtained if the free energy is known by the equation $\mu = \delta F / \delta n$. The free energy will be chosen in the form suggested by the Landau model:

$$F[n] = \int d\vec{r}\left(\frac{K}{2}(\vec{\nabla}n)^2 + f(n) + nV\right) \qquad (3)$$

The term $\frac{K}{2}(\vec{\nabla}n)^2$ describes the energy due to inhomogeneity. The additional energy acquired by excitons in the non-uniform potential, is taken into account by the term $nV$.

Substituting Eqs. (2) and (3) into Eq. (1) we obtain

$$\frac{\partial n}{\partial t} = \vec{\nabla}\left(M\vec{\nabla}\left(-K\Delta n + \frac{df}{dn} + V\right)\right) + G - \frac{n}{\tau}. \qquad (4)$$

Later on we expand the functions $f$ and $M$ in the power series of $n$. The phase transition occurs in the vicinity of the minimum of $f(n)$. We expand the $f$ in series of n up to the forth power which is sufficient to introduce the minimum.

$$f(n) = \kappa Tn(\ln n - 1) + \frac{a}{2}n^2 + \frac{b}{3}n^3 + \frac{c}{4}n^4 \qquad (5)$$

where $a$, $b$ and $c$ are phenomenological parameters. Since the $M$ is the smooth function of n we take into account the first nonzero term in its expansion in powers of $n$. In this case $M = Dn/(\kappa T)$, where $D$ is the diffusion coefficient of the free exciton in the well.

The first term in Eq. (5) gives the typical expression $D\Delta n$ in Eq. (1) in the limit of the low density. In the case of the high exciton density, which is the range of our interest, this term is not important. The others terms in Eq. (5) are related to the exciton-exciton interaction. For a fixed value of the electric field across the double quantum well and increasing pumping at first the term $\frac{a}{2}n^2$ plays the major role. In this case the main contribution into exciton-exciton interaction comes from the dipole-dipole repulsion, which is the reason of the blue shift of the exciton emission line observed with increasing pumping. Therefore, the value of the parameter $a$ is positive ($a > 0$). As the exciton density growth further, the distance between excitons decreases and the exchange and the van der Waals interactions begin to contribute. These interactions may give a negative contribution into the free energy. Additionally, in order to obtain a stable state solution at $n \to \infty$ the free energy given by Eq. (5) requires $c > 0$. As we assume that a condensed phase exists, the free energy should have a minimum at a certain value of $n$. Therefore, the condition $b < 0$ should apply.

The free energy may be expanded in the series of $(n - n_m)$, where $n_m$ is the position of the minimum of $f(n)$. In this case $f(n) = f(n_m) + a_1(n - n_m)^2 + b_1(n - n_m)^3 + c_1(n - n_m)^4$. In such form the free energy was used in the papers [15,17] for the explanation of structures observed in the works [4,6].

For the parameters of the system when the exciton density is close to $n_m$, both approaches give similar results. We shall use the free energy in the form given by Eq. (5).

Let us perform the normalization choosing $l_0 = (K/a)^{1/2}$ as the unity length, $n_0 = (a/c)^{1/2}$ as the unity exciton density, $\tau_0 = (\kappa T K c^{1/2})/(D a^{5/2})$ as the unity time and introducing new notations: $\tilde{\tau} = \tau/\tau_0$, $\tilde{G} = G c K \kappa T/(D a^3)$, $\tilde{b} = b/\sqrt{ac}$, $\tilde{D} = \kappa T c^{1/2}/a^{3/2}$, $\tilde{V} = V c^{1/2}/a^{3/2}$. The dimensionless exciton lifetime depends on many parameters: the exciton lifetime, which can be controlled by the size of the interwell barrier, parameters of the condensed phase, the diffusion coefficient and others. The estimations show that the dimensionless exciton lifetime may vary in a wide range $(10 \div 10^4)$.

Later on we shall omit the symbol ~.

In the dimensionless units Eq. (1) for the exciton density may be rewritten in the following form

$$\frac{\partial n}{\partial t} = D\Delta n + div(n\vec{\nabla}(-\Delta n + n + bn^2 + cn^3 + V)) + G - n/\tau \qquad (6)$$

As shown in [18] in the case of the attractive interaction between excitons the uniform distribution of the exciton density becomes unstable at a certain critical value of the pumping.

Summarizing our approach, we will use the nonlinear Eq. (6) for the exciton density instead of the nonlinear Gross-Pitayevsky equation typically used for the wave function of the excitonic condensate. The rationale for this is in the fact that the exciton free path caused by a disorder is of order of the distance between excitons and much less than the order of the typical size of a non-homogeneity $(((1 \div 10) m\mu)$, which appears when the emission of patterns [4,6] are formed. As the result the wave function of the condensate loses its coherence.

## 3. The additional potential for excitons in the vicinity of a slot in an electrode and the exciton distribution at the low intensity of excitation

Let us consider a semiconductor double quantum well heterostructure sandwiched between two metallic electrodes: the top electrode containing a transparent slot with the width of $2b$, and the bottom electrode covering the whole lower surface of the sample (Figure 1).

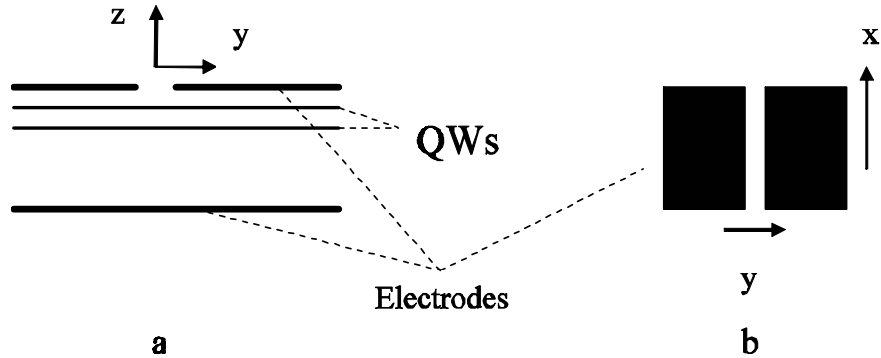

Figure 1. The arrangement of quantum wells and electrodes in semiconductors in the case of a slot in the top electrode: a) side view, b) view from above.

In order to solve the equation describing the density distribution of excitons created under the slot by light, we first determine the energy of an exciton localized in the double well as a function of coordinates relative to the slot in the electrode. Let us choose the $X$ axis along the slot and the $Z$ axis perpendicular to the electrodes. When the voltage is applied to the electrodes the indirect excitons acquire an additional energy $V = -p_z E_z$, where $p_z$ is the dipole moment of an exciton, directed along the $Z$ axis in the strong electric field. The electric field under the slot is not uniform. In order to determine its strength it is necessary to solve the Laplace equation for the potential with the following boundary conditions: the potential should be constant on both electrodes and the potential difference between the electrodes must be such that the field between the electrodes is equal to $E_0$ far from the slot. To this end we use the method of the solution of Laplace equation in ellipsoidal coordinates (see [28]). We have used the solution presented in [28] for the problem of determination of the field created by a grounded metallic plate with a slot placed in the uniform external electric field. In application to our problem this solution does not satisfy the condition of the constant potential on the lower electrode. However, the corrections to the potential induced by the presence of the slot decrease with distance from the slot, and, therefore, are small near the lower electrode for $b \ll L$, where $L$ is the distance between the electrodes. For this reason, we assume that the plane of the well is located much closer to the upper electrode than to the lower electrode. Moreover, using the solution of the problem considered in [28], where the medium is the same on both sides of the electrode with the slot, we assume that the upper electrode (the electrode with the slot) is buried in the semiconductor. A well with high conductivity may serve as an example of such an electrode [6]. Using these approximations the additional potential energy created by the presence of the slot in electrode may be presented in the following form

$$V(y,z) = \frac{V_0}{2}\left( (\sqrt{1+b^2/\xi(y,z)} - 1) - \frac{b^2 z^2}{2\xi(y,z)^2 \sqrt{1+b^2/\xi(y,z)}} \left(1 + \frac{y^2+z^2+b^2}{\sqrt{(y^2+z^2-b^2)^2+4b^2 z^2}}\right)\right) \quad (7)$$

where

$$\xi(y,z) = \frac{1}{2}(y^2+z^2-b^2+\sqrt{(y^2+z^2-b^2)^2+4b^2 z^2}) \quad (8)$$

$V_0 = -p_z E_0$ is the shift of the exciton band caused by the electric field far from the slot.

In our problem the coordinate $z$ determines the distance of the quantum well from the upper electrode. The potential $V(y,z)$ created by the slot is the function of the ratios $y/b$ and $z/b$. Figure 2 depicts the typical behavior of the potential as a function of $y$ for a certain set of the parameters of a particular geometry of the slot.

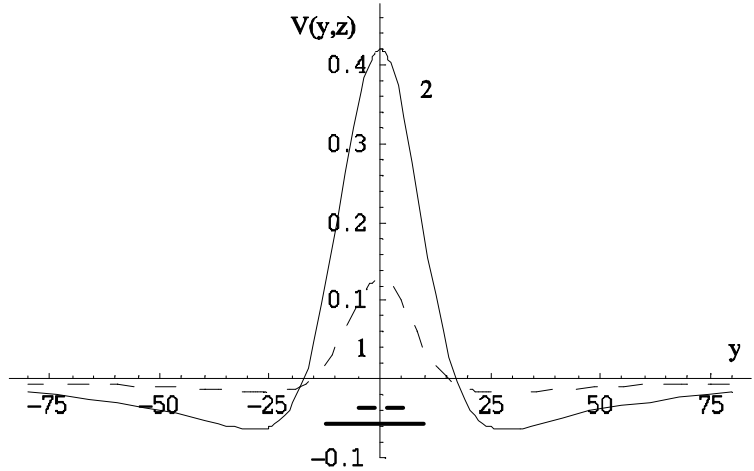

Figure 2. The dependence of the potential created by the slot on the distance from the center of the slot in the plane of the quantum well for the following values of the dimensionless parameters: $b_1 = -2.23$, $\tau = 100$, $D = 0.2$, $V_0 = -5$, $z = -15$, $b = 5$ for the curve 1 and $b = 10$ for the curve 2.

Because the electric field in the regions of the quantum well under the slot is less than the field far from the slot the additional potential for excitons is positive. Therefore, the slot creates a potential hump for an exciton in the center of the slot. But in a certain vicinity of the borders of the slot the potential has a small minimum with a negative value. It appears due to the rearrangement of charges on the conductive electrode in the vicinity to edges of the slot. The depth of the minimum increases with increasing the width of the slot ($b$) and becomes constant in the limit $z \ll b$.

Let us analyze the distribution of the exciton density at the irradiation of a low intensity when the interaction between excitons is not important. In the case of the irradiation of the system by a laser with a wide beam the excitons are created in the quantum well in the region with the width $2b$, so $G(\vec{r}) = G(y) = G$ for and $G(\vec{r}) = G(y) = 0$ for $y < -b$, $y > b$. This region is shown in Fig. 2 by the dashed and solid lines for the curves 1 and 2, respectively. After the excitation the excitons slide down in the field of the inhomogeneous external potential $V(y,z)$. But due to the finite value of the lifetime excitons penetrating under the electrode cannot move far from the slot. As the result, the exciton density distribution has a maximum under the rim of the window. The shape of the spatial distribution of the exciton density and the position of the maximum depend on the width of the slot, the exciton lifetime, the diffusion coefficient and the form of the additional potential. This distribution, obtained from the solution of Eq. (6) in the linear approximation with respect to $n$, is presented in Figure 3.

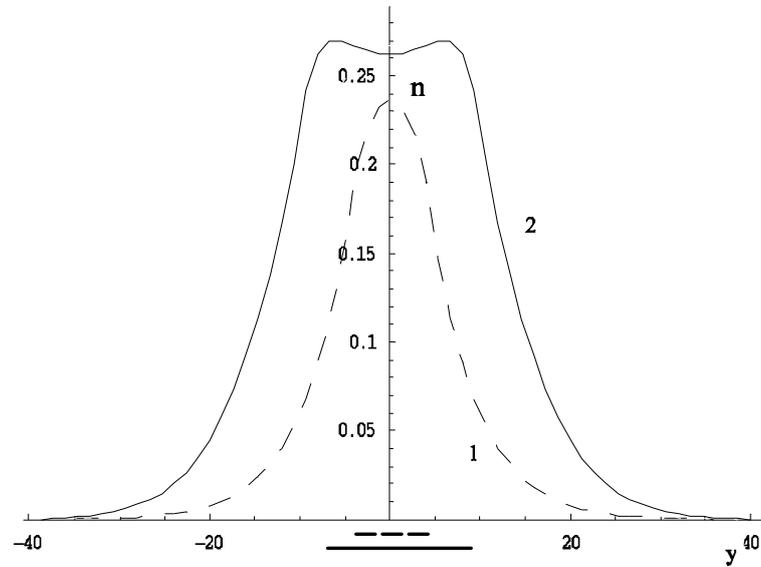

Figure 3. The dependence of the exciton density at the low intensity excitation on the distance from the center of the slot for the following values of the dimensionless parameters: $b_1 = -2.23$, $\tau = 100$, $D = 0.2$, $V_0 = -5$, $G = 0.004$, $z = -15$, $b = 5$ for the curve 1, $b = 10$ for the curve 2

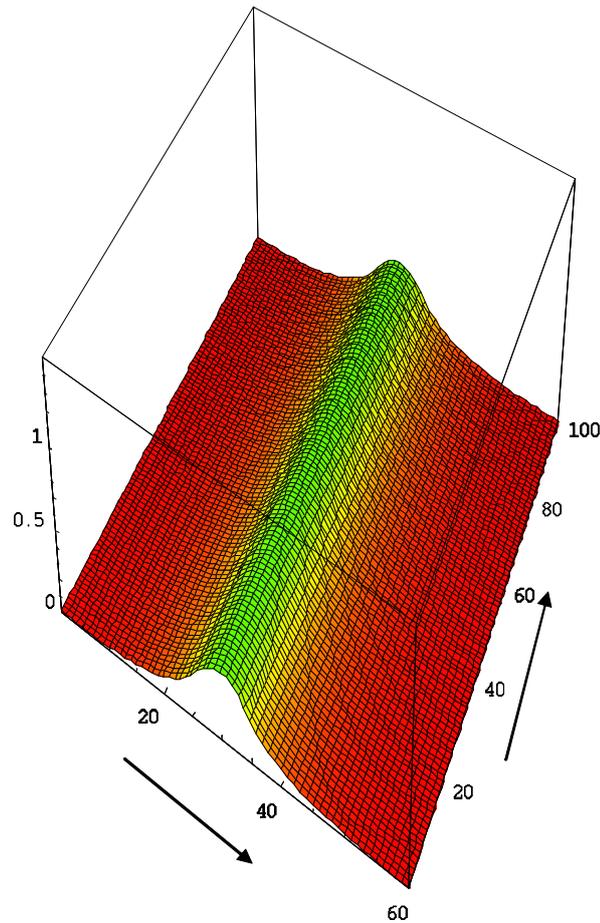

Figure 4. The spatial distribution of the exciton density at the low intensity excitation for the following values of the dimensionless parameters: $b_1 = -2.23$, $\tau = 100$, $D = 0.2$, $V = -5$, $G = 0.007$, $z = -15$, $b = 7$.

For a narrow slot the exciton density has a maximum in the center of the slot (Figure 3, curve 1 and Figure 4). With increasing the width the maxima of the exciton density are developed at the edges of the slot (see curve 2 in Figure 3). In general case the positions of the maxima do not coincide with the positions of the minima of the additional potential (see Figure 2). For example, according to Figure 3, the maximum density for the curves 2 takes place at $y = 8$, while for the corresponding value of $b$ the minimum of the potential $V$ in Figure 2 occurs at $y = 27$. The position of the maximum distribution function depends significantly on the exciton lifetime.

## 4. The structure formation at high exciton density

In the case of the high exciton density the nonlinear Eq. (6) was solved numerically for a strip in the XY plane extending beyond the slot on both sides. The following boundary conditions were chosen: the normal projection of the exciton flux at the boundaries of the strip was set to be equal to zero. The size of the strip was chosen sufficiently large to make the results insensitive to it. The obtained results are as follows.

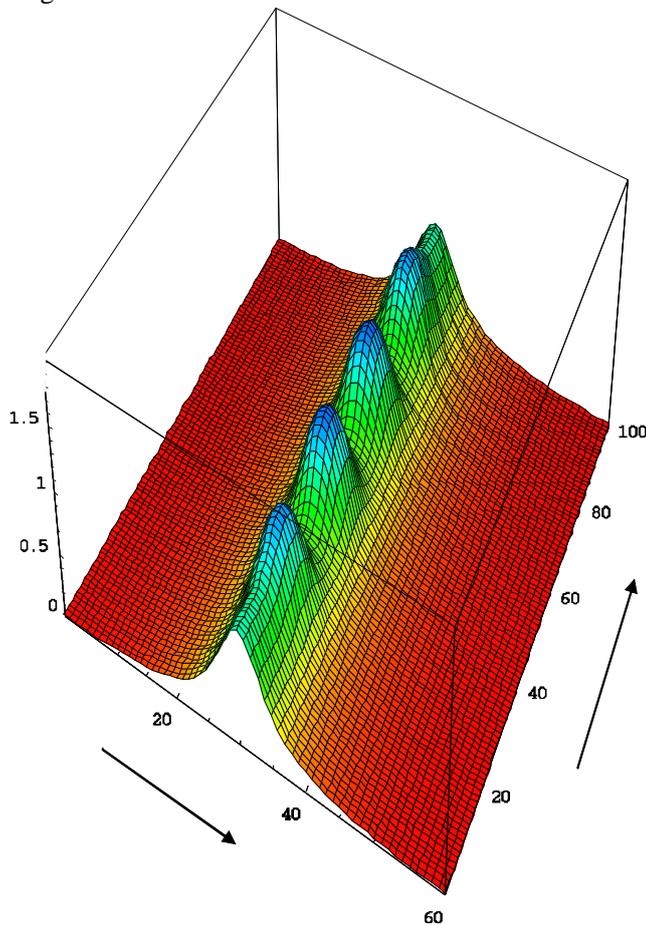

Figure5. The spatial distribution of the exciton density at the low intensity excitation for the following values of the dimensionless parameters: $b_1 = -2.23$, $\tau = 100$, $D = 0.2$, $V_0 = -5$, $G = 0.009$, $z = -15$, $b = 7$.

According to the Figure 3 and Figure 4 for a small value of $b$ and for the low intensity irradiation the exciton density has a maximum in the region directly under the center of the slot. With increasing the pumping the uniform distribution of the exciton density along the slot becomes unstable and a periodical structure arises (Figure 5). Islands of the condensed phase of excitons alternate with regions of the gas phase. The threshold value of the exciton generation rate, at which the periodical structure appears, increases with decreasing the width of the slot.

As the pumping increases further the periodical structure transforms into a continuous distribution with a high value of the exciton density in the center (Figure 6) extending along the whole length of the slot. For the system presented in Figure 6 the density in the center exceeds the value created by the pumping directly. The reason for that is the attractive interaction between excitons which gathers the excitons in the center despite the fact that the external potential pulls them away from the center. This phenomenon should be observed in experiment as a spatial narrowing of the strip of emission from the middle of the slot.

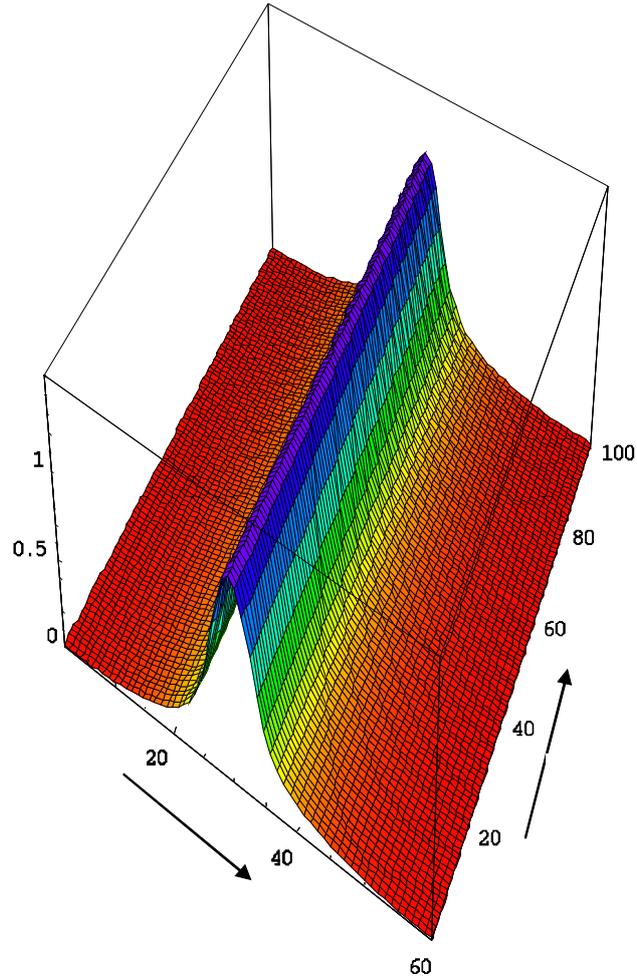

Figure 6. The spatial distribution of the exciton density at the low intensity excitation for $G = 0.01$. The other parameters are the same as in Figure 5.

With increasing the width of the slot the maxima of the exciton density are developed at the edges of the slot (see curve 2 in Figure 3) and the pattern of the islands undergoes interesting changes. In this case two parallel chains of islands localized at the opposite sides of the slot arise instead of a single chain in the center (Figure 7). The positions of islands in the chains are shifted by a half of the chain period with respect to each other.

The results of the paper are presented in dimensionless units which is a convenient way to perform theoretical calculations. It is useful to provide an example of the results for a particular system expressed in dimentinal units making realistic suggestions about the values of the parameters that enter the expression for the free energy. The values of the real parameters characterizing indirect excitons, such as the lifetime and the diffusion coefficient, depend strongly on the structure of the double quantum well. We shall take typical values for them given in literature. As the presented theory is phenomenological, the parameters in the expression for the free energy are unknown and should be extracted from experimental data. The parameter $a$ in Eq. (5) may be estimated from the blue shift of the exciton levels with increasing the exciton density in the range lower than the critical density of the condensation. For the values of other parameters we rely on the analysis of the experiment [6] carried out in the paper [17]. For example, we suggest the following values for the parameters for the system: $\tau = 10^{-8} s, D = 10 cm^2 s^{-1}$, $an_0 = 2 \cdot 10^{-3} eV$, $n_0 = 3.2 \cdot 10^{10} cm^{-2}$, $l_0 = 0.7 \mu m$. The dimension values of still other parameters $(K, c, b_1)$ can be expressed through the mentioned above. For such values of the parameters of the exciton system the results presented in Figure 5 correspond to the following dimension values for the system with the slot: the width of the slot $2b = 10 \mu m$, the distance from the double quantum well to the electrode $10.5 \mu m$, the applied bias $V_0 = -10^{-2} eV$, the pumping $G = 2.9 \cdot 10^{18} cm^{-2} s^{-1}$. The period of superlattice in Figure 5 equals $14\,\mu m$ and the maximal value of exciton density in the chain equals $3.6 \cdot 10^{10}\,cm^{-2}$. The double chain of the condensed phase islands is observed in Figure 7 at $2b = 14\,\mu m$.

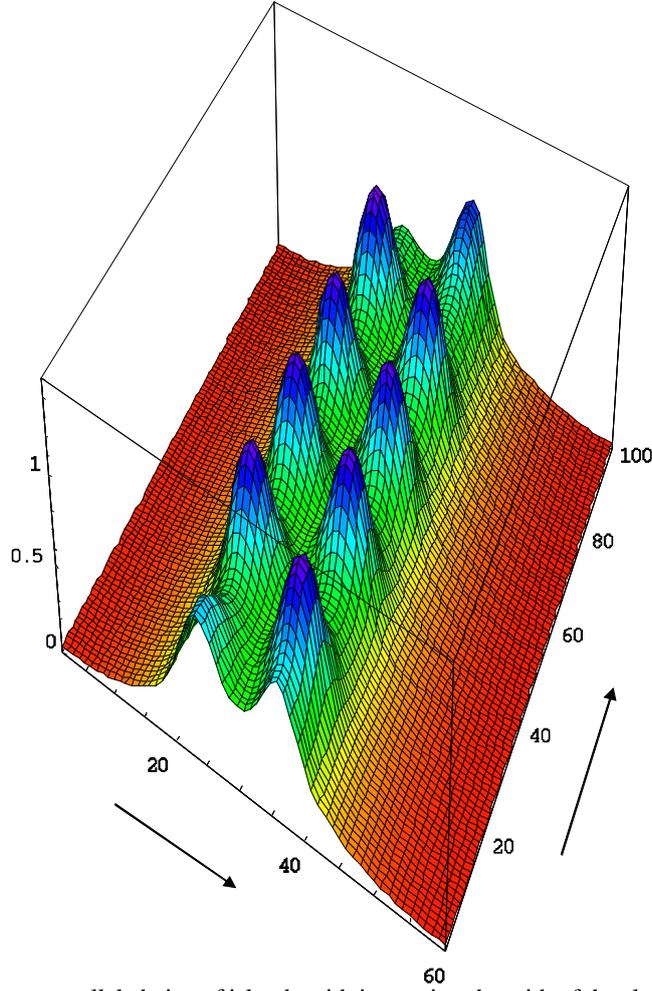

Figure 7. The formation of two parallel chains of islands with increasing the with of the slot. The parameters of the system are as follows: $b_1 = -2.23$, $\tau = 100$, $D = 0.2$, $V_0 = -5$, $G = 0.009$, $z = -15$, $b = 10$.

## 5. Moving islands of the condensed phase of excitons in quantum wells. Excitonic analog of Gunn effect

Let us suggest a system in which additionally to the slot there exists an external potential with a gradient along the $x$ axis: $V_h(x, y, z) = V(y, z) + \delta V x$, where $V(y, z)$ is the potential determined earlier by Eq. (7). The additional potential $\delta V x$ may be created by the inhomogeneous pressure.

It is simple to show that in the potential $\delta V x$ there exists an auto-wave solution of Eq. (6) for the exciton density, i.e. the solution with the time and space coordinates connected by the relation $\xi = x - vt$ that takes the form $n_h(x, y, z) = n(x - vt, y, z)$, where $n(x, y, z)$ is the steady-state solution for the exciton density distribution obtained above, and $v = -\delta V$ is the velocity of the auto-wave. In this manner the structure determined by Eq. (6) moves with the velocity $v$. So, the exciton density is a periodical function of the coordinate drifting with a certain velocity along the slot. And the exciton density changes periodically as a function of time in a certain point of the space.

Let us make some estimations. In the dimensional units the velocity is equal to $v = -D\delta V/(\kappa T)$. In the case of $D = 10\,cm^2/s$, $T = 2K$, $\delta V = 10^{-3}\,eV/(100\,\mu m)$ we obtain $v = 5.7 \cdot 10^3$ cm/s. It should be noted that every island of the condensed phase contains many excitons (more than 1000) and, therefore, its motion is accompanied with the external field driven transfer of the energy of order of !0³-10⁴ eV.

The considered effect resembles the Gunn effect known in the semiconductor physics. Both effects are nonlinear. In the Gunn effect, a fluctuation of the charge distribution (a domain) moves in the crystal, while in the considered case it is the island of the condensed phase of excitons that drifts along the slot. The inhomogeneous potential plays the role similar to the role of the electric field in the Gunn effect. But, in contrast to the Gunn effect, where the number of particles is conserved, in the considered case the excitons are constantly created and decay.

**6. Structures in the system with periodical external potential**

Let us consider the case when the external potential is periodic function in some direction. We shall approximate the potential by the function

$$V(x) = V_0 \cos^2(2\pi k y) \qquad (9)$$

So the external potential is periodic function along $x$ direction and does not depend on $y$. We also suggest that distribution of excitons which is created by a laser in double quantum well may be described by the formula

$$G(\vec{r}) = \frac{P_0}{2\pi R^2} \exp\left(-\frac{x^2 + y^2}{2R^2}\right) \qquad (10)$$

Resently such type potential was realized in double quantum well and investigation of localization and delocalization of excitons in minimum of this potential was fulfilled [29].

In present work we shall not analyze the exciton density distribution for potential (9) in detail, we produce only the solution of the equation (6) for a small plate in the case of three case of the laser intensity.

At the small laser intensity excitons occupy the regions in the minimums of potential (9) and uniform distribution of exciton density along the axis y is realized (Figure 8).

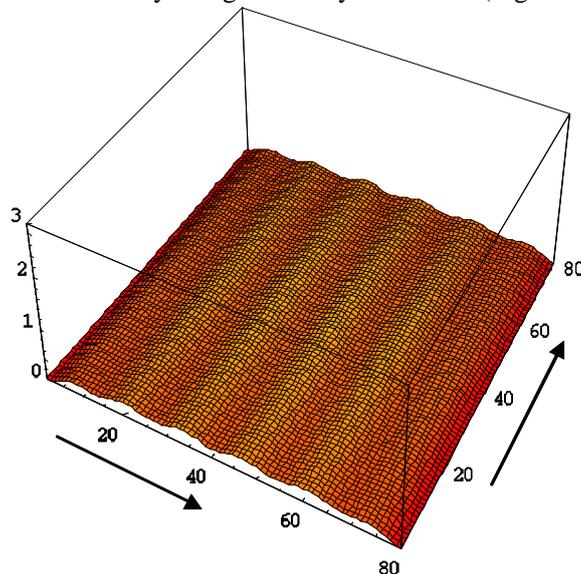

Figure 8. Distribution of exciton density in quantum well in presence of periodical external potential at low intensity of pumping. The parameters of the system are following: $\tau = 100$, $D = 0.1$, $b_1 = -2.3$, $V_0 = 0.1$, $k = 3$, $R = 60$, $P_0 = 60$

With increasing the pumping the structure along the axis $y$ appears (Fig9).

The structure occurs as the result of the condensed phase appearance : the regions in minima of potential consist of islands of condensed phases and the regions of gas phases.. At high pumping the sizes of the islands increases and exciton distribution becomes uniform along the y-axis (Figure 10).

The detailed picture of exciton density distribution depends on a depth of the potential, a period of external potential and other parameters of the system.

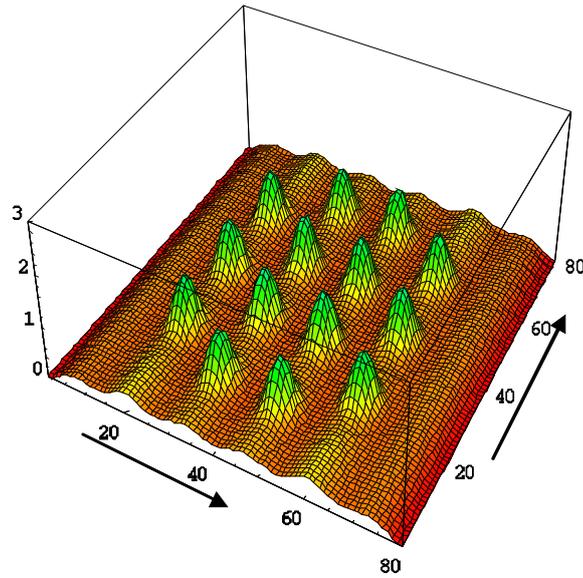

Figure 9. Distribution of exciton density in quantum well in presence of periodical external potential.. $P_0 = 90$ The other parameters are the same as in Figure 8.

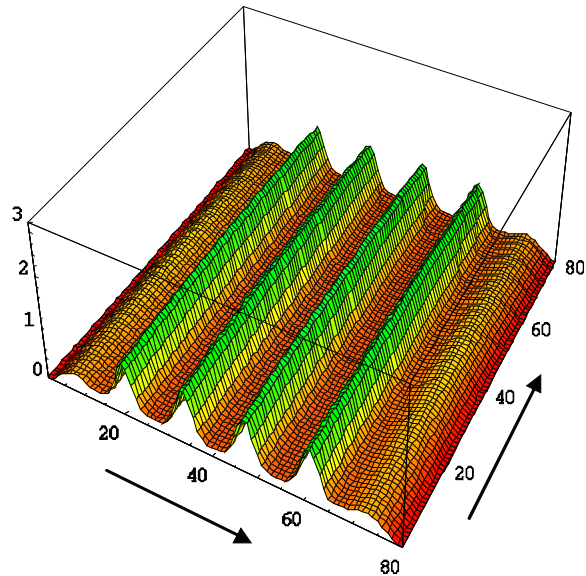

Figure 10. Distribution of exciton density in quantum well in presence of periodical external potential. $P_0 = 120$. The other parameters are the sameas in Figure 8.

### 6. Conclusions

The paper studied the distribution of the density of indirect excitons in a double quantum well in the spatial region under a transparent slot in one of metallic electrodes. The excitons are created by the light irradiation through the slot. It is assumed that there exists a condensed phase of excitons which is described phenomenologically. For the determination of the exciton density in the quantum well the model of the spinodal decomposition of the phase transitions theory generalized for the case of unstable particles is used. The structures of the exciton density distribution in the vicinity of the slot are studied depending on the pumping, the width of the slot, the distance between the quantum well and the electrode. It is shown that at a certain value of the pumping a periodical chain of islands of exciton condensed phases arises. With increasing the width of the slot the chain splits into two chains shifted by a half period of the chain with respect to each other. If an additional potential with a linear dependence on coordinate along the slot is applied the chain moves along the slot with a constant velocity. During such drift the exciton density in a certain point of the system is a periodical function of time. The period of the oscillations is also determined by the parameters of the potential. The effect should manifest itself in the form of a periodical variation of the intensity of the light emitted from a certain region under the slot. In the case of one-dimensional periodical potential along one axis at some values of pumping the islands of condensed phases arise along both axis.

The patterns arise at parameters at which the exciton condensed and gaseous phases coexistent. The risen of the appearance of pattern is non-equilibrium state of the system connected with the finite value of exciton lifetime.

At presented phenomenological consideration the microscopic model of condensed phase is not determined. It is suggested in the presented model that the condensed phase arises due to attractive interaction between particles. The condensed phase may be not only exciton liquid , but also the electron-hole liquid. In last case the value $n$ describes the density of electron-hole pairs. In principle the islands of condensed phase may be in superfluid state. For example, in classical case of liquid helium the liquid state arises firstly from gas phase due to attractive van der Waals interaction between atoms and then at further decreasing of temperature the superfluid state appears.

The experimental study of the suggested setup gives the possibility to build periodical structures controlled by light.

The author would like to thank Dr. Goliney I.Yu. for fruitful discussions.

**References**
[1] Lozovik Yu. E., Yudson V.I. 1976 *Sov. Phys JETP* **44** 389
[2] Zinets O.S., V.I. Sugakov, A.D. Suprun  1976 *Sov.Phys.Semicond* **10** 422
[3] Zinets O.S., V.I. Sugakov, A.D. Suprun 1978 *Pis 'ma Zh. Tekhn. Phys*. **1** 61 (in Russian)
 [4] Butov L. V., Gossard A. C., and Chemla D. S. 2002 *Nature* **418** 751
Butov L. V., Levitov L. S., Mintsev A. V.,. Simons B. D., Gossard A. C., and Chemla D .S 2004 *Phys. Rev. Lett*. **92** 117404
[5] Snoke D., Denev S., Liu Y., Pfeiffer L., and West K. 2002 *Nature* **418**, 754
Snoke D., Liu Y., Denev S., Pfeiffer L., and. West K. 2003 *Solid State Commun*. **127** 187
Rapaport R., Chen G., Snoke D., Simon S. H., Pfeiffer L., West K., Liu Y., and Denev S. 2004 *Phys. Rev. Lett*. **92** 117405 .
[6] Gorbunov A. V. and Timofeev V.B. 2006 *Pis'ma Zh.Eksp.Teor. Phys*. **83** 178-784 [2006 *JETP Lett*. **83**, 146]
Gorbunov A .V. and Timofeev V. B. 2006 *Usp. Fiz. Nauk* **176** 651
Gorbunov A. V., Timofeev V. B. 2006 *Pis'ma Zh.Eksp.Teo. Phys*.**84**, 390  [2006 *JETP Lett*. **84** 329 ]
Timofeev V. B., Gorbunov A. V., Larionov A. V. 2007 *J. Phys. Condens. Matter* **19** 29529
[7] Levitov L. S., Simons B. D.,. Butov L. V. 2005 *Phys.Rev. Lett*. **94** 176404
[8] Liu C. S,.Luo H. G., Wu W. C. J. 2006 J. *Phys.:Condens. Matter*.**18** 9659
[9] Paraskevov A., Khabarova T. V. 2007 *Phys. Lett*.,A **368** 151
[10] Saptsov R..B. 2007 *JETP Lett*. **105** 566
[11] Sugakov V. I. 1986 *Fiz. Tverd. Tela* (Leningrad) **21** 562 -[1986 *Sov. Phys. Solid State* **21**,332]
[12] Sugakov V.I. 2002 *Phase Transitions* **75** 953
[13] Sugakov V. I. 2004 *Fiz. Tverd. Tela* (Leningrad) **46** 1458-[2004 *Sov. Phys. Solid State* **46**,1496]
[14] Sugakov V. I. 2005 *Solid State Commun*. **134**, 63 [arXiv:cond-mat/040739]
[15] Chernyuk A. A. and Sugakov V. I. 2006 *Phys. Rev*. B **74** 085303 [arXiv:cond-mat/0508282]
[16] Sugakov V. I. 2007 *Phys. Rev*. B **76** 115303
[17] Sugakov V. I.. Chernyuk A. A. 2007 *JETP LETTERS* **85** 570
[18] Dremin A.A., Timofeev V. B., Larionov A. V., Hvam J and. Soerensen K 2002
*JETP Letters* **76** 526
[19] Lozovik Yu. E., Berman O. L. 1996 *Pis'ma v Zh. Eksp. Teor. Fiz*. **64** 526
[20] Tan M. Y. I ,Drumord N. D. , and Needs I. 2005 *Phys.Rev*. B **71** 033303 .
[21] Schindler Ch., Zimmermann R. 2008 *Phys. Rev*. B **78** 045313
[22] Lee R.M., Drummond N.D., and Needs R.J. 2009 *Phys.Rev*. B **79** 125308
[23] Hammack A.T., Griswold M., Butov L.V., Smallwood L.E., Ivanov A.L., Gossard A.C. 2006 *Phys. Rev. Lett*. **96** 227402
[24] Voros Z., Snoke D.W., Pfeiffer L., and West K. 2006 *Phys. Rev. Lett*. **97** 016803
[25] Gartner A., Prechtel L., Schuh D., Holleitner A.W., and Kotthaus J.P. 2007 *Phys.Rev*. B **76** 0805304
[26] Kubo R., Yokota M., Nakajima S. 1957 *Journ.Phys, Soc. Japan* **12** 1203
[27] Zubarev D. N. 1974 *Nonequilibrium statistical thermodynamics*( New York Plenum Consaltants Bureau)
[28] Landau L. D. and. Lifshits E. M. 1982 *Course of Theoretical Physics*. Vol.8.
 *Electrodynamics of Continuous Media*. 2nd Ed. (Moskow Nauka) [1984 Oxford Pergamon]
[29] M. Remeika, J.C. Graves, A.T. Hammack, A.D. Meyertholen, M.M. Folger, L.V. Butov, M. Hanson, A.C. Gossard. 2009 Phys.Rev. Lett. **102** 186803 [*arXiv:0901*.1349vl]